# RF CAVITIES FOR THE MUON AND NEUTRINO FACTORY COLLABORATION STUDY *

A. Moretti, N. Holtkamp, T. Jurgens, Z. Qian and V. Wu, FNAL, Batavia, IL 60510, USA

## Abstract

A multi-laboratory collaboration is studying the feasibility of building a muon collider, the first phase of which maybe a neutrino factory. The phase space occupied by the muons is very large and needs to be cooled several orders of magnitude for either machine, 100,000 to 1 million for the collider and ten to 100 for the factory. Ionization cooling is the baseline method for muon cooling. This scheme uses hydrogen absorbers and rf re-acceleration in a long series of magnetic focusing channels to cool the muons. At Fermilab two rf cavity types are under study to provide the required cooling rf re-acceleration. A 805 MHz high gradient cavity for the collider and a 201 MHz high gradient cavity for the neutrino factory. The 805 MHz cavity currently under going cold testing is a non-periodic pi-mode cavity with the iris openings shaped to follow the contour of the beam. The 201 MHz cavity uses hollow thin metal tubes over the beam aperture to terminate the field in a pillbox type mode to increase its shunt impedance. This is possible because muons have little interactions with thin metal membranes. Details of these cavities and cold measurement data will be presented.

## 1 INTRODUCTION

An international collaborative study of muon colliders and neutrino factories has been going on for a number of years [1, 2, 3]. The lead laboratories for this study are BNL, CERN, Fermilab and LBL. A collider or neutrino factory for high-energy research needs a large number of muons to produce a high luminosity beam. Because of the muons short lifetime, they need to be transported quickly through the accelerator complex. The muons are produced from pions decays off of a proton beam hitting a high-Z target in a solenoidal magnetic decay channel. The muons, thus, produced occupy a very large 6-demensional phase space which must be reduced (cooled) quickly by several orders of magnitude to meet the luminosity requirements. Ionization cooling has been chosen as the cooling technique. In this technique muons lose transverse and longitudinal momentum as they pass through a low-Z absorbing material. The longitudinal momentum is then restored by rf re-acceleration in large aperture rf cavities. The process is repeated numerous times to reduce the 6-demensional phase space of the muons sufficiently for acceptance by the accelerator complex and meet its luminosity requirements.

*Work supported by the US Dept. of Energy, contract DE-AC02-76CH0-3000.

Currently, at Fermilab and LBL high gradient, high shunt impedance large beam aperture rf cavities are being studied at 201 and 805 MHz [4]. Accelerating gradients of 15 MV/m and 30 MV/m for 201 and 805 MHz respectively are required for the most favored scenarios. LBL is studying pill-box type rf cavities with 125 micron beryllium windows over the aperture, due to their higher shunt impedance and low radio of peak surface field to accelerating field. Fermilab is studying a 201 MHz cavity with thin hollow beryllium or aluminum tubes over the aperture. The tubes terminate the aperture electric fields in a pillbox type mode and increase its shunt impedance towards that of a true pill-box cavity. At Fermilab, also, a 805 MHz open cell cavity has been designed and a cold model has been built and tested. To increase its shunt impedance the iris openings have been dimensioned to follow the beam's contour as it passes through the cavity.

## 2 GRIDDED 201 MHZ CAVITY DESIGN

The 201 MHz gridded cavity is bellow shaped to increase its shunt impedance and has a set of crossed (gridded) hollow thin walled low-Z metallic tubes covering the bean aperture, Fig.1. The tubes can be easily forced gassed cooled, a great advantage over 125 micron Be window covering the beam aperture. In this design the tubes are made of aluminum 4 cm in diameter, 125 microns thick in the middle and 500 microns at its ends. The cavity has a 0.60 m major radius, a length of 0.64 m and beam aperture of 0.64 m. Current mechanical and electrical designs limit the Be window aperture design to 0.38 m. When connected to neighboring cavities, the cavities are separately driven with a phase advance of pi per cavity. Other phase advances are possible because the grids were designed to minimize the coupling between neighboring cavities.

The computer program MAFIA was used to optimize the design of the cavity. The number of tubes and their diameters were varied to maximize the shunt impedance, reduce the peak surface electric field, minimize material intercepting the beam and the coupling between neighboring cavities with the beam aperture set at 0.64 m. Following the above criteria, four vertical and four horizontal tubes, 4 cm in diameter resulted in the most satisfactory cavity design. The cavity, Fig. 1, has a $Q_o$ of 63,000, shunt impedance of 32.0 MOhm/m and requires 4.5 MW to achieve a accelerating gradient of 15 MV/m. The peak surface electric field at this gradient is 25 MV/m, an acceptable 1.7 times the Kilpatrick Limit.

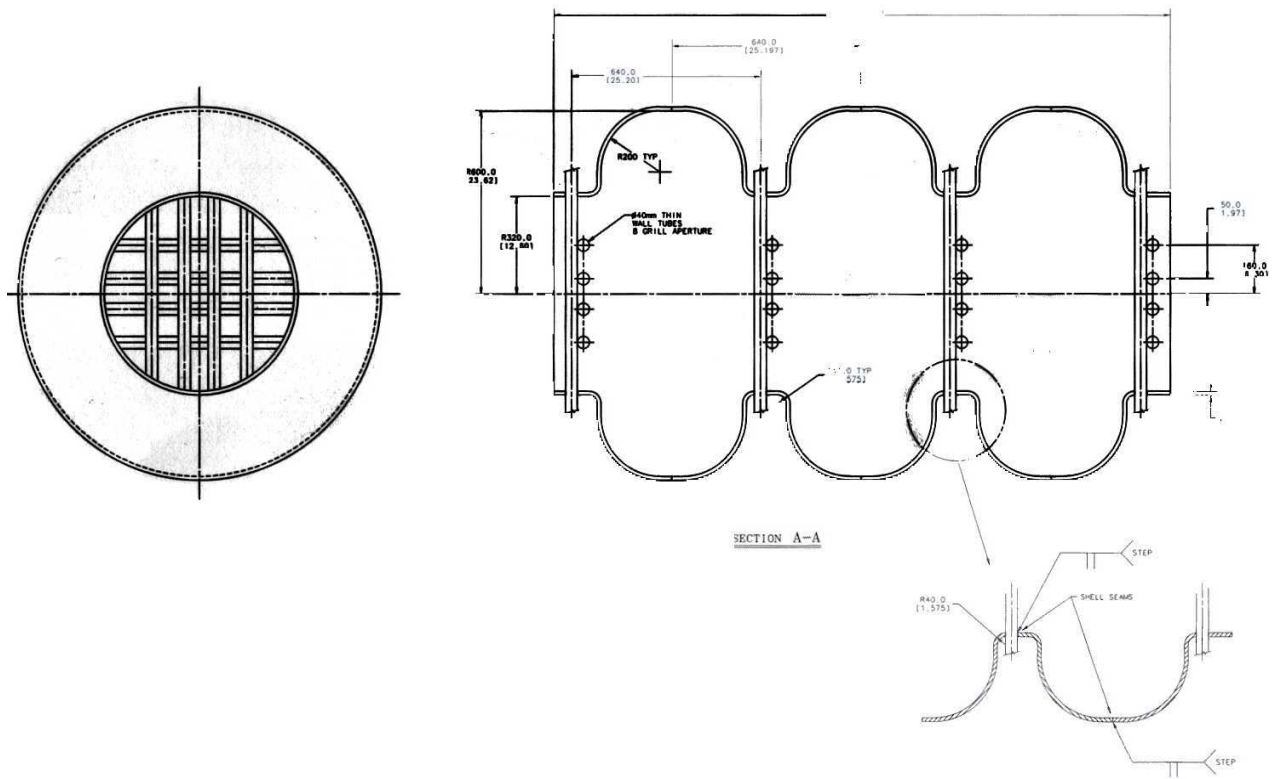

Figure 1: Aperture Tube Layout

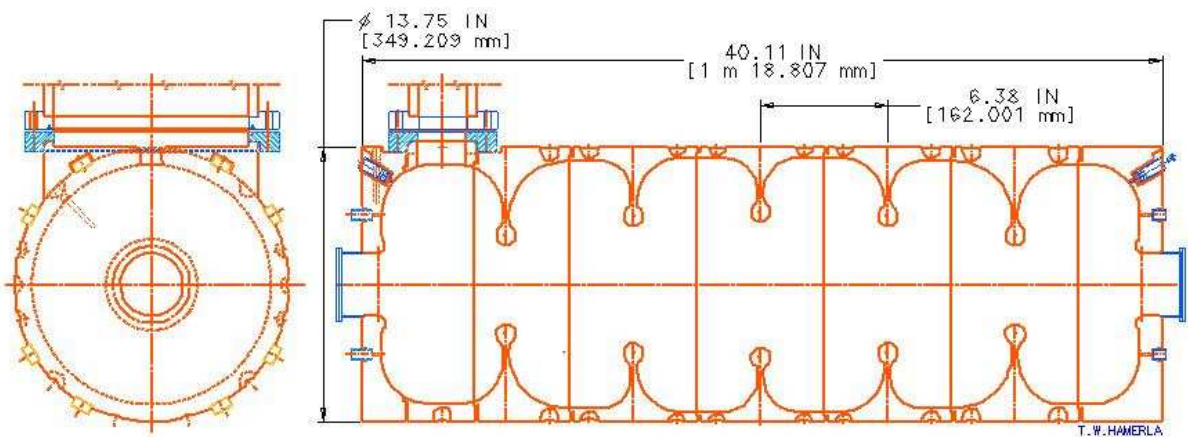

Figure 2: 805 MHz iris loaded cavity with a beam envelope matched aperture.

## 3 OPEN CELL 805 MHZ CAVITY DESIGN

The 805 MHz cavity is an iris-loaded structure with the aperture of the iris dimensioned to follow the five-sigma contour of the beam, Fig. 2. This allows the design to maximize the shunt impedance without material in the beam path. This may increase cooling channel efficiency. However, cooling simulations have shown little if any improvement when compared to 125 Be window design. The design does eliminate the difficult mechanical and rf electrical heating problems of the Be window design. The beam aperture in the middle of the cavity is 0.16 m and at the ends 0.08 m.

The computer programs Mafia and Superfish were used to optimize the design. The criteria of the design was to maximize shunt impedance while obtaining nearly equal and reasonable peak surface electric fields on all the cavity irises. The cavity, shown in Fig.2, has a Qo of

35,600, a shunt impedance of 33.5 Mohm/m and requires 27.7 MW for a accelerating gradient of 30 MV/m. The peak surface electric field is 77 MV/m, 2.9 times the Kilpatrick limit. This might be acceptable for the required short cooling pulse length of 210 microseconds. A high power copper vacuum cavity is currently under construction and breakdown studies are planned in a Fermilab test facility under construction.

A full-scale aluminum model of the cavity has been built. The model was built to test the accuracy of the computer calculation and the machining accuracy of the parts. The machining accuracy called for was +/- 13 microns. Measurements on the model were very good. The measurements agreed with calculations to within 5 microns. Bead-pull measurements of the field profile were in agreement with calculations to within 5 %. The model was further used to determine the size of the critical coupling slot.

## 4 HIGH POWER RF COUPLER DESIGN

Mafia 3D time domain and 2D eigenmode solvers are used for the coupler simulations [5]. The model consists of the first two cells of the six cell cavity with a rectangular waveguide (the height is one half of the standard WR975 waveguide height) attached to the outer wall of the first cell (see Figure 2). Energy coupling between the waveguide and the cavity is through a rectangular slot. The height of the slot is chosen to be that of the waveguide, in order to minimize the ratio of the maximum coupler voltage to waveguide voltage, while the depth is the outer wall thickness of the cell. The width is varied to achieve critical coupling. All corners of the coupling slot's cross section are rounded to a radius of 7 mm. To simulate the total wall loss in cell 3 through 6 in the actual cavity, the conductivity of the second cell is adjusted to produce the loss. The conductivity is determined using the 2D eigenmode solver where the wall loss of each cell can be calculated; hence the conductivity. In the simulation, it is important that cell 1 and 2 are in tune and have the correct relative energy distributions.

The coupling coefficient ($\beta$) calculation employs the energy method [5] in which two-time domain runs are needed. In the first time domain run, each cell is tuned separately to 805 MHz and to have the right energy distribution. After tuning, the two-cell structure is excited by a monochromatic dipole signal located inside the second cell. The 3D electric and magnetic fields are recorded at four carefully chosen time steps. From these fields, the power loss at the cavity wall and the power flow out of the cavity into the waveguide are calculated. The coupling coefficient is computed as the ratio of external power over wall loss. A low power test is performed on a full-scale (six-cell) aluminum model to check the simulations. For a set of coupler dimensions that is close to critical coupling ($\beta = 1$), the measured $\beta$ is 0.976. The simulation result is 1.000. Finally, the coupling slot dimensions for critical coupling are determined to be (height, depth, width) = (6.2, 2.2, 8.2) cm.

## 5 CAVITY RESEARCH STATUS

Measurements of the frequency, field profile and coupling slot size on the aluminum model were in excellent agreement with Mafia and Superfish calculations. A 805 MHz copper high power test cavity is under construction. Electric field breakdown and vacuum conditioning studies are to take place in a high power test facility currently under construction at Fermilab. Design studies of high gradient 201 MHZ cavities are currently in progress at Fermilab and LBL. The goal of these studies is to produce several high power prototype cavities in the next two years. A test facility at Fermilab is currently in the early design stage.